\documentclass[reprint,aps,pra,showkeys, onecolumn, notitlepage]{revtex4-1}
\usepackage[pdftex]{graphicx}
\usepackage[dvipsnames]{xcolor}
\usepackage{color}
\usepackage{amsmath}
\usepackage{amssymb}
\usepackage{dcolumn}
\usepackage{bbold}
\usepackage{braket}
\usepackage{lipsum} 
\usepackage{caption}
\usepackage{subcaption}
\usepackage{amsthm}
\usepackage{setspace}
\usepackage{hyperref}
\usepackage[T1]{fontenc}
\definecolor{darkblue}{RGB}{0,0,149}
\hypersetup{
    colorlinks=true,
    linkcolor=red,
    filecolor=darkblue,
    urlcolor=darkblue,
    citecolor=Green,
    linktocpage=true,
}
\usepackage{url}

\makeatletter

\def\p@section{}
\def\p@subsection{}
\makeatother

\begin{document}

\title{Second-quantized Unruh-DeWitt detectors\\ and their quantum reference frame transformations}

\author{Flaminia Giacomini}%
 \email{fgiacomini@perimeterinstitute.ca}
\affiliation{Perimeter Institute for Theoretical Physics, 31 Caroline St. N, Waterloo, Ontario, N2L 2Y5, Canada
}

\author{Achim Kempf}
\email{akempf@perimeterinstitute.ca}
\affiliation{Department of Applied Mathematics, University of Waterloo, Waterloo, ON N2L 3G1, Canada}

\affiliation{Department of Physics, University of Waterloo, Waterloo, ON N2L 3G1, Canada	
}

\affiliation{Institute for Quantum Computing, University of Waterloo, Waterloo, ON N2L 3G1, Canada}
\affiliation{Perimeter Institute for Theoretical Physics, 31 Caroline St. N, Waterloo, Ontario, N2L 2Y5, Canada
}

\begin{abstract}
   
We generalize the Unruh-DeWitt detector model to second quantization. We illustrate this model by applying it to an excited particle in a superposition of relativistic velocities. We calculate, to first order, how its decay depends on whether its superposition of velocities is coherent or incoherent. Further, we generalize the framework of quantum reference frames to allow transformations to the rest frames of second-quantized Unruh DeWitt detectors. As an application, we show how to transform into the rest frame of a decaying particle that, in the laboratory frame, is in a linear superposition of relativistically differing velocities.  
$$$$
\end{abstract}

\maketitle

\section{Introduction}

Some of the most fundamental properties of the light-matter interaction do not depend significantly on its spinorial, vectorial and gauge theoretic aspects. These properties include, for example, the Hawking and Unruh effects~\cite{hawking1975particle, unruh1976notes, unruh2005universality, crispino2008unruh}, the possibility to harvest entanglement from the vacuum~\cite{valentini1991non, olson2012extraction, sabin2012extracting, lorek2014extraction, henderson2018harvesting, simidzija2018general, stritzelberger2021entanglement}, and the ability to establish a quantum channel via quantum fields~\cite{cliche2010relativistic, firstorderisgood, simidzija2020transmission, jonsson2020communication}. For this reason, such basic properties of the light-matter interaction are often investigated using the simpler Unruh-DeWitt (UDW) model~\cite{BirrellDavies}. In the UDW model, the quantized electromagnetic field is simplified to a quantized scalar field. Further, matter such as an atom is modeled as a localized first-quantized 2-level system, a so-called Unruh-DeWitt  detector. The motion of UDW detectors has conventionally been described either as a classical path~\cite{BirrellDavies, Hummer:2015xaa}, which can also be smeared~\cite{Martin-Martinez:2020pss, Martin-Martinez:2020lul, Perche:2021uws}, or more recently through a first-quantized Schr\"odinger wave function \cite{Stritzelberger:2019gnc}. Using either classical paths or Schr\"odinger wave functions to describe the motion of UDW detectors has the drawback that either quantum effects or relativistic effects are missed. 
Therefore, in Sec.~\ref{sec2}, we generalize the Unruh-DeWitt detector model to second quantization. To this end, we model UDW detectors as excitations in the one-particle sector of a quantum field. As we then show in Sec.~\ref{sec3}, this allows one, for example, to study how the decay of an excited UDW detector in a superposition of relativistic velocities depends on whether the superposition of velocities is coherent or incoherent. 

Further, in Sec.\ref{sec4}, we generalize the formalism of quantum reference frames (QRF) to allow the transformation to rest frames that are defined by second-quantized particles, such as second-quantized UDW detectors. The ability to transform into rest frames is an important tool because there are observables and processes that are most naturally defined in rest frames. For example, the use of QRF methods was shown to give an operational identification of a relativistic spin operator. Such an operational identification is useful to encode quantum information in a massive special-relativistic particle moving in a quantum superposition of relativistic velocities~\cite{giacomini2019relativistic, streiter2020relativistic}.
In the case here, as we show in Sec.~\ref{sec5}, the application of QRF methods allows one to calculate aspects of the decay of an unstable particle in its rest frame, even if, from the perspective of the lab frame, the unstable particle is in a coherent linear position of relativistically differing velocities. To this end, we here make use of methods from Ref.~\cite{giacomini2019relativistic,streiter2020relativistic}. These methods allow us to transform to the rest frame of a decaying UDW detector which moves in a superposition of (relativistic) velocities from the perspective of the laboratory, define and calculate the relevant observables there, and then transform back to the laboratory frame, in which measurements are made.

Tools of QRF  transformations have been developed since 1967 in the fields of both quantum information~\cite{aharonov1, aharonov3, brs_review, spekkens_resource, palmer_changing, katz2015mesoscopic, smith_quantumrf, poulin_dynamics, busch_relational_1, busch_relational_2, busch_relational_3, angelo_1} and quantum gravity~\cite{dewitt1967quantum, Kuchar:1990vy, Brown:1994py, Brown:1995fj, rovelli_quantum, Rovelli:2004tv, Dittrich:2004cb, Tambornino:2011vg, rovelli_relational}. Here, we make use, in particular, of the formalism for QRFs introduced in Ref.~\cite{Giacomini:2017zju}, see also related work~\cite{perspective1, perspective2, hoehn2018switch, hoehn2019switching, hoehn2019trinity, giacomini2019relativistic, hoehn2020equivalence, hardy2020implementation, yang2020switching, castro2020quantum, de2020quantum, krumm2020, mikusch2021transformation, tuziemski2020decoherence, giacomini2020einstein, ballesteros2020group, streiter2020relativistic, giacomini2021spacetime, hoehn2021quantum, hoehn2021internal, giacomini2021quantum, castro2021relative, de2021perspective, cepollaro2021quantum, de2021entanglement, de2021falling}. The core of this formalism is a generalisation of standard reference frame transformations to quantum superpositions of reference frame transformations. Such superpositions of reference frame transformations naturally arise when a quantum system that serves to define a rest frame is in a coherent superposition of velocities, as can be the case here with the second-quantized UDW detectors.
In this case, for example, proper time is then subject to a quantum superposition of special relativistic time dilations~\cite{Smith:2019imm, Grochowski:2020mru, giacomini2021spacetime}. This phenomenon is described here in Sec.~\ref{sec4} via the \emph{quantum superpositions of Lorentz boosts}, which transform from the laboratory frame to the rest frame of the second-quantized UDW detector and viceversa. Using these QRF transformations, as we show in Sec.~\ref{sec5}, if an UDW detector starts in its excited state, observables related to its decay can be calculated (here to first order in the coupling) in its rest frame and then transformed into the lab frame.  

\section{Introducing second-quantized Unruh-DeWitt detectors}

\label{sec2}

\subsection{Unruh-DeWitt detectors}
In quantum field theory, particles can be defined as excitations of mode oscillators. Alternatively, and this will be the approach that we follow here, a particle can be defined as what a particle detector detects. Here, a particle detector can be any quantum system with some internal energy levels that interacts with the quantum field. We then say that the detector has detected a particle  if the detector system has transitioned from the ground state into an excited state. 
For example, an atom can serve in this way as a detector system for particles of the electromagnetic field. 

The notion of a small particle detector system is very versatile and is often used not only for the description of the detection of particles but also of their emission and other more exotic processes. In many circumstances, the details of the detector system and the field can thereby be neglected to some extent, see, e.g., \cite{udwisgood}. Examples are studies of the Unruh and Hawking effects~\cite{hawking1975particle, unruh1976notes, unruh2005universality, crispino2008unruh}, quantum communication through quantum fields~\cite{cliche2010relativistic, firstorderisgood, simidzija2020transmission, jonsson2020communication}, quantum energy teleportation~\cite{hotta2010controlled, verdon2016asymptotically} and the extraction of entanglement from the vacuum~\cite{valentini1991non, olson2012extraction, sabin2012extracting, lorek2014extraction, henderson2018harvesting, simidzija2018general, stritzelberger2021entanglement}. 
   
In such circumstances, the detector system can often be idealized by considering only two of its energy levels, by giving the detector system a classical trajectory, and by modeling it as coupled to a scalar field. Such an idealized, so-called Unruh-DeWitt detector is a qubit with an energy gap $\Omega$, i.e., with a free Hamiltonian $H_D=(\Omega/2) \sigma_3$, travelling on a prescribed classical trajectory $x(t)$ and coupled to the quantum field through an interaction Hamiltonian of the form:
\begin{equation}
H_{int}(t) = \lambda(t) \sigma_1 \int d^3x'~\chi(x(t)-x')~\hat{\Phi}(x') \label{hint}
\end{equation}
Here, $\lambda(t)$ describes the coupling strength, which is allowed to depend on time to enable the description of switching. The smearing function $\chi$ describes the spatial extent of the UDW detector. The first Pauli matrix $\sigma_1$ enters because it obeys $\sigma_1=\sigma_-+\sigma_+$, so that the interaction Hamiltonian allows the field to excite and de-excite the detector. Conversely, since the field can be expanded in creation and annihilation operators, $\hat{\Phi}(k)\propto (u(k)a_k+ u^*(k)a_k^\dagger)$, Eq.~\eqref{hint} allows the detector to create and destroy field quanta. 

In the interaction Hamiltonian, the resulting terms of the form $\sigma_+a$ and $\sigma_-a^\dagger$ are the so called rotating wave terms and they describe the detector's excitation with simultaneous absorption and the de-excitation of the detector with simultaneous emission respectively. These are the only contributing terms when the UDW detector-field system is closed, so that its energy is conserved. 

The so-called counter-rotating terms, i.e., the terms of the form $\sigma_+a^\dagger$ and $\sigma_-a$, contribute when time-translation invariance is broken, e.g., through the switching of the interaction, through measurement at a finite time, through acceleration or through gravity. In particular, the counter-rotating terms of the form $\sigma_+a^\dagger$ enable the `detection' of particles without the de-excitation of a mode oscillator. Mathematically, this is related to the occurrence of nonzero $\beta$-coefficients in calculations of these processes through Bogolubov transformations. Physically, this leads, e.g., to the particle creation predicted in the Unruh and Hawking effects. In recent work, \cite{soda2021accelerationinduced}, it was shown for the Unruh effect that fine-tuned stimulation makes it possible to strongly amplify the counter-rotating terms while arbitrarily strongly suppressing rotating wave terms, in a phenomenon of acceleration-induced transparency.

The starting point for us here will be a recent generalization of the UDW detector model that treats the center of mass of the detector system quantum mechanically~\cite{Stritzelberger:2019gnc,sudhir2021unruh}. This model was used, for example, to show that virtual velocities of UDW detectors, e.g., in the process of quantum delocalization, can lead to a new type of Cherenkov effect when the virtual velocities exceed the maximum propagation speed of field perturbations in a medium. 
The generalized UDW detector model was also used to show, for example, that entanglement harvesting into UDW detectors' internal degrees of freedom is strongly obstructed by recoil, since some of the harvested entanglement tends to go into the UDW detectors' external degrees of freedom~\cite{stritzelberger2021entanglement}.

A key limitation of this model of UDW detectors has been, however, that the description of the quantum mechanics of the center of mass of an UDW detector is nonrelativistic.
Here, we therefore generalize this UDW detector model to second quantization. The price to pay is that we lose the perfect localizability of UDW detectors of prior UDW detector models. (This is a small price to pay since perfect localizability is an over-idealization in the sense that there cannot be a position operator in second quantization.)
The benefit is that the new second-quantized model of UDW detectors includes their quantized external degrees of freedom fully relativistically. This allows one, for example, to consider UDW detectors which are in a superposition of velocities that differ relativistically. This then enables correspondingly generalized, and therefore more realistic studies, for example, of entanglement harvesting or quantum communication through quantum fields. 

\subsection{Generalization to second-quantized UDW detectors}
\label{sec1.2}
In order to obtain a covariant model of UDW detectors that possess quantized external degrees of freedom, we begin with the observation that, as an UDW detector system such as an atom gets excited from the ground state energy level $E_0$ to an excited state energy level $E_1$, its rest mass correspondingly increases~\cite{zych_interferometric, pikovski, CastroRuizE2303}. This motivates modeling Unruh-DeWitt detectors in second quantization as the quanta of a massive scalar field $\psi$ which consists of two subspecies that differ by their rest masses. For a process described in second quantization leading to a detector model, see Refs.~\cite{Perche:2021ycr, bruno}.

For ease of terminology, we will sometimes refer to the UDW detector as an atom and the scalar field as the electromagnetic field. For simplicity, we let our UDW detectors interact, as usual, with a massless field $\Phi(x)$ that is scalar. For now, we work in an inertial reference frame and we choose the interaction picture. We work in units where $\hbar=c=1$. The mode expansions then read:
\begin{align}
\psi(x) = \begin{pmatrix}
  \psi_{0}(x) \\
   \psi_{1}(x)  \\
   \end{pmatrix}, \qquad   & \psi_a (x) = \frac{1}{(2\pi)^{3/2}}\int \frac{d^3 p}{\sqrt{2\omega_{p,a}^D}} \left( b^\dagger_{p,a}\, e^{i p\cdot x} +  b_{p,a}\, e^{-i p\cdot x}\right), \qquad a=0,1\\
    & \Phi(x) = \frac{1}{(2\pi)^{3/2}}\int \frac{d^3 k}{\sqrt{2|\vec{k}|}} \left( a^\dagger_k\, e^{i k\cdot x} +  a_k\, e^{-i k\cdot x}\right)
\end{align}
Here, $\omega_{p,a}^D := \sqrt{|\vec{p}|^2 + \left(m_D + E_a \right)^2}$ denotes the dispersion relation of an UDW detector particle, which we may call D, that is either excited ($a=1$) or in the ground state ($a=0$). The rest mass of an UDW detector in the ground state is denoted by $m_D$. For any four-vector, $v$, $\vec{v}$ indicates a three-vector. Further, $p \cdot x = \omega_{p,a}^D \Delta t- \vec{p}\cdot \vec{x}$ and $k \cdot x = |\vec{k}| \Delta t- \vec{k}\cdot \vec{x}$. The free and interaction Hamiltonians read: 
\begin{align} \label{HamFree_Sch_UDW}
        \hat{H}_{free} =& \sum_{a=0,1} \int d^3 p \,\omega_{p,a}^D \hat{b}_{p,a}^\dagger \hat{b}_{p,a} + \int d^3 k \, |k|\, \hat{a}_{\vec{k}}^{\dagger} \hat{a}_{\vec{k}}^{\vphantom \dagger},\\
    \hat{H}_{int} =& \lambda \int d^3 x \psi^\dagger(x) \hat{\sigma}_1 \psi(x) \Phi (x), \qquad \hat{\sigma}_1 = \begin{pmatrix} 0 & 1\\ 1 & 0 \end{pmatrix}  \label{HamInt_Sch_UDW}
\end{align}

This interaction Hamiltonian contains terms which describe excitation/de-excitation and emission/absorption as well as now also processes that change the number of UDW detectors, as well as the scattering of radiation off of UDW detectors. The  emission of a photon from an atom, for example, is now modeled as the simultaneous annihilation of an excited (i.e., higher rest mass) UDW detector, the creation of an unexcited (i.e., lower rest mass) UDW detector and the creation of a photon. The corresponding terms in the interaction Hamiltonian are of the form $b^\dagger_0 a^\dagger b_1$ with their hermitean conjugates describing absorption.   

In most of what follows, for simplicity, we will focus on the resonant phenomena of emission and absorption, i.e., we only keep the rotating wave terms and neglect the counter-rotating terms.
 Further, we drop scattering terms (namely, those that do not change the internal state of the detector) and those terms in the interaction Hamiltonian that describe processes that change the number of UDW detectors. Therefore, we obtain the following resonant interaction Hamiltonian, $H_{int}^{(res)}$:
\begin{eqnarray}
            \hat{H}^{(res)}_{int} &=&  \frac{\lambda}{(2\pi)^{9/2}}\int d^3 x \frac{d^3 k}{\sqrt{2|k|}} \frac{d^3 p}{\sqrt{2\omega_{p,0}^D}} \frac{d^3 p'}{\sqrt{2\omega_{p', 1}^D}} e^{i (p +k - p')\cdot x}\, b^\dagger_{p,0} a^\dagger_k b_{p',1} + h.c.\\
            &=&   \frac{\lambda}{(2\pi)^{3/2}} \int \frac{d^3 k}{\sqrt{2|k|}} \frac{d^3 p}{\sqrt{2\omega_{p,0}^D}} \frac{1}{\sqrt{2\omega_{p+k, 1}^D}} \,  e^{i (\omega_{p,0} +|k| - \omega_{p+k,1})t} \,b^\dagger_{p,0} a^\dagger_k b_{p+k,1} + h.c. \label{Hint}
    \end{eqnarray}
For example, to describe spontaneous emissions, we can choose the initial state $\ket{\Psi(0)}$ to be a state with one UDW detector prepared in the excited state ($a=1$), and with the $\Phi$ field initially in its vacuum state:
\begin{equation} 
    \ket{\Psi(0)} = \int \frac{d^3p}{\sqrt{2\omega_{p,1}^D}} ~\phi_0(p) b^\dagger_{p,1}\ket{0}_D \ket{0}_E.
\end{equation}
Here, $\vert 0\rangle_E$ is the vacuum of the $\Phi$ field that stands in, for example, for the electromagnetic field, $\vert 0\rangle_D$ is the vacuum of the UDW detector field and $\phi_0(p)$ describes the wavepacket in momentum space of one excited UDW detector D. To first order perturbation in $\lambda$, at some final time $t$, the system evolved into a coherent superposition of the initial state and a state in which the decay of the detector and the emission of a photon has occurred:    
\begin{equation} \label{eq:FullStateinA}
    \ket{\Psi(t)} = \left[ \mathbb{1} -i \int_{0}^{t} dt'\hat{H}^{(res)}_{int}(t') \right] \ket{\Psi(0)} =  \ket{\Psi(0)} + \lambda \ket{\psi(t)}.
\end{equation}
Here: 
\begin{equation} \label{eq:FinalStateA}
    \ket{\psi(t)} =\frac{-i}{(2\pi)^{3/2}} \int_{0}^{t} dt'\int \frac{d^3p}{\sqrt{2\omega_{p, 0}^D}} \frac{d^3k}{\sqrt{2|k|}} \frac{1}{2\omega_{p+k,1}^D} \phi_0(p+k) e^{it'(\omega_{p,0} +|k| - \omega_{p+k,1})} b^\dagger_{p,0}\ket{0}_D a^\dagger_k\ket{0}_E.
\end{equation}

\subsection{Recovering a prior model in the first quantized nonrelativistic limit}
Before moving on to an application, we now show that one can recover the description of  UDW detectors with first quantized external degrees of freedom of \cite{Stritzelberger:2019gnc, stritzelberger2021entanglement,sudhir2021unruh} as a special case. 
The interaction Hamiltonian of this model reads, in the Schr\"odinger picture:
\begin{equation}
    \hat{H}'_{int} = \lambda ~\hat{\sigma}_1\otimes{\Phi}(\hat{\vec{x}})=\lambda\int d^3x~\vert \vec{x}\rangle\langle\vec{x}\vert\otimes\hat{\sigma}_1\otimes{\Phi}(\vec{x})\label{Hintp}
\end{equation}
Here, the second expression is defined, using functional calculus, in terms of the third expression, see also \cite{unruhwald}. 
Our aim now is to show that the interaction Hamiltonian $\hat{H}_{int}'$ of this model arises as a special case from $\hat{H}_{int}$. To this end, we again drop from $\hat{H}_{int}$ the terms that would describe scattering or changes in the number of UDW detectors, but this time we keep the counter-rotating terms, to obtain from $\hat{H}_{int}$ the simpler Hamiltonian 
$\hat{H}_{int}^{(rec)}$:
\begin{equation}
\hat{H}_{int}^{(rec)}(t) = \lambda \sum_{a\neq a'} \int d^3 x ~\Phi (x) \int \frac{d^3 p}{\sqrt{2\omega_{p,a}^D}} \frac{d^3 p'}{\sqrt{2\omega_{p',a'}^D}} e^{i p\cdot x} b_{p,a}^\dagger b_{p',a'}e^{-i p\,'\cdot x} ~ + ~h.c. 
\end{equation}
We now restrict the new second-quantized model to the one-particle sector of the UDW detector field to obtain an effectively first quantized model of UDW detectors, and we take the nonrelativistic limit. 
First, when restricting the action 
of $\hat{H}_{int}^{(rec)}$ 
to the vacuum and one-particle sector ${\cal{H}}^D_1$ of the Hilbert space of the UDW detector field, we can also write:
\begin{equation}
\hat{H}_{int}^{(rec)}(t)\vert_{{\cal{H}}^D_1} = \lambda \sum_{a\neq a'} \int d^3 x ~\Phi (x) \int \frac{d^3 p}{\sqrt{2\omega_{p,a}^D}} \frac{d^3 p'}{\sqrt{2\omega_{p',a'}^D}} e^{i p\cdot x} b_{p,a}^\dagger \ket{0}_D \bra{0}_D b_{p',a'}e^{-i p\,'\cdot x} ~ +~ h.c.
\end{equation}
By defining the states
\begin{equation}
    \ket{{x}\,, a}_D := \int \frac{d^3 p}{\sqrt{2\omega_{p,a}^D}}\, e^{it\omega_{p,a}}e^{-i \vec{p}\cdot \vec{x}} b^\dagger_{p,a} \ket{0}_D,\label{npes}
\end{equation}
the interaction Hamiltonian can be written in a form which almost matches that of $H_{int}'$ of Eq.~\eqref{Hintp}: 
\begin{equation} \label{eq:nonrelHam}
    \hat{H}_{int}^{(rec)}(t)\vert_{{\cal{H}}^D_1}
    = \lambda \sum_{a\neq a'} \int d^3 x \ket{{x}\,, a}_D \bra{{x}\,, a'} \otimes\Phi({x}) 
\end{equation}
The problem is that the states $\ket{x\,, a}_D$ are not position eigenstates. This is because covariance demands that the integral measure in Eq.~\eqref{npes} contains a $\vec{p}$ dependence through $\omega_{p,a}^D$. This implies that the states $\ket{x\,, a}_D$ are not exactly the Fourier transforms of plane waves in momentum space. This is an expression of the fact that there are no position operators in relativistic quantum field theory. For a recent detailed investigation of this and related phenomena, see \cite{mariajason}. However, in the nonrelativistic regime, where the rest mass of the UDW detector dominates over the occurring momenta, $\omega_{p,a}^D$ is approximately constant. This implies that, in the nonrelativistic regime, Eq.~\eqref{npes} becomes a Fourier transform and therefore, while the states $\ket{{x}\,, a}_D$ live in the one particle sector of the UDW detector field, they then behave approximately as the position eigenstates of first quantization. In the nonrelativistic regime, with the one-particle sector playing the role of first quantization, Eq.~\eqref{eq:nonrelHam} therefore recovers, in the interaction picture, the nonrelativistic first quantized UDW detector model of \cite{Stritzelberger:2019gnc} whose interaction Hamiltonian in the Schr\"odinger picture is given by Eq.~\eqref{Hintp}.

\section{Application: Decay of a particle with coherent vs. incoherent superposition of velocities.}
\label{sec3}
We now return to the second-quantized model of UDW detectors of Sec.~\ref{sec1.2}. As a simple example, we study the decay  of an excited UDW detector as a function of time to first order perturbation theory. Using the framework of second-quantized UDW detectors above, we can now model the excited UDW to start in arbitrary quantum states of motion. In particular, we can now calculate and compare the cases where the excited UDW detector is prepared in a state which represents either a coherent or an incoherent superposition of two initial velocities, even if the two velocities differ relativistically. 

\subsection{Excited UDW detector in coherent versus incoherent superposition of velocities}

Let us pick two arbitrary momenta $p_1,p_2$. The normalized state $\ket{\Psi_i(0)}$ that describes at the initial time $t=0$ an excited UDW detector of momentum $p_i$ and the field $\Phi$ in the vacuum state reads, for $i=1,2$:
\begin{equation} 
    \ket{\Psi_i(0)} =   
    b^\dagger_{p_i,1}\ket{0}_D \ket{0}_E
\end{equation}
The time-evolved state with the Hamiltonian of Eq.~\eqref{Hint} is, to first order perturbation in $\lambda$: 
\begin{eqnarray} 
    \ket{\Psi_i(t)}&=&\ket{\Psi_i(0)} + \lambda \ket{\psi_i(t)}  \\
    &=&\ket{\Psi_i(0)} - i \int_0^t dt'~ H_{int}^{(res)}(t')~ 
    b^\dagger_{p_i,1}\ket{0}_D \ket{0}_E\\
    &=&\ket{\Psi_i(0)} -i\int_o^t dt'~\lambda \int \frac{d^3 k}{\sqrt{2|k|}} \frac{1}{\sqrt{2\omega_{(p_i-k),0}^D}} \frac{(2\pi)^{3/2}}{\sqrt{2\omega_{p_i, 1}^D}} \,  e^{it'(\omega_{(p_i-k),0} +|k| - \omega_{p_i,1})} \,b^\dagger_{p_i-k,0} a^\dagger_k   \ket{0}_D \ket{0}_E  \label{1storder}
\end{eqnarray}
For later reference, notice that we here defined vectors
\begin{equation}
   \ket{\psi_i(t)} := - \frac{i}{\lambda} \int_0^t dt'~ H_{int}^{(res)}(t')~\ket{\Psi_i(0)}
\end{equation}
which are not normalized. In fact,
$    s_i(t):=\braket{\psi_i(t)|\psi_i(t)}$
obeys $s_i\rightarrow 0$ as $t\rightarrow 0$.
Throughout, as the need arises, we will tacitly assume infrared regularization such as a box regularization. 
Let us now consider the two cases of the UDW detector starting either in an equal coherent or equal incoherent superposition of these two momenta. In the interaction picture, we have for the coherent superposition: 
\begin{eqnarray}
    \rho_c(t) &=& \left(\frac{1}{\sqrt{2}}\sum_{i=1,2}\ket{\Psi_i(t)}\right)\left(\frac{1}{\sqrt{2}}\sum_{j=1,2}\bra{\Psi_j(t)}\right)\\
    \label{coh}
    &=& 
    \frac{1}{2}\left(\sum_{i=1,2}\ket{\Psi_i(0)}+\lambda \ket{\psi_i(t)}\right)
    \left(\sum_{j=1,2}\bra{\Psi_j(0)}+\lambda \bra{\psi_j(t)}\right)
\end{eqnarray}
For the incoherent superposition we have:
\begin{eqnarray}
    \rho_{ic}(t) &=& \frac{1}{2}\sum_{i=1,2}\ket{\Psi_i(t)}\bra{\Psi_i(t)}\\
    &=& \frac{1}{2}\sum_{i=1,2}\left(\ket{\Psi_i(0)}+\lambda \ket{\psi_i(t)}\right)
    \left(\bra{\Psi_i(0)}+\lambda \bra{\psi_i(t)}\right)
    \label{incoh}
\end{eqnarray}
Our calculations of the quantum state $\ket{\psi_i(t)}$ are valid up to the first order in $\lambda$, hence the density operators that we obtain when substituting Eq.~\eqref{1storder} into Eq.~\eqref{coh} and Eq.~\eqref{incoh} are also valid only up to the first order in $\lambda$. To this order, we can discard the second order terms in $\lambda$ contained in $\rho_{c}(t)$ and $\rho_{ic}(t)$. (To make the density operator valid to order $\lambda^2$, additional terms, obtained from calculating $\ket{\Psi_i(t)}$ to second order, would be needed.) As we will now show, subtle nontrivial effects can already be seen to first order in $\lambda$. 

\subsection{Observables}
Our aim is to determine under which circumstances the presence of coherence terms (i.e., of off-diagonal terms in momentum basis) in $\rho_c(t)$ versus their absence in $\rho_{ic}(t)$ might affect observables during the decay process.

As the simplest possible candidates for such observables, let us consider basic binary observables, $\hat{Q}$, i.e., observables that are rank one projectors (modulo continuum normalization). To this end, we define the vector
\begin{equation}
   \ket{q(t)}:= \sum_{i=1,2}\left(\alpha_i \ket{\Psi_i(0)} + \beta_i \ket{\psi_i(t)}\right)
\end{equation} 
where the $\alpha_i$ and $\beta_i$ are arbitrary complex coefficients such that $\ket{q(t)}$ is normalized at the time $t$ when we want to perform a measurement. We can then define a binary observable as the rank one projector $\hat{Q}(t) := \ket{q(t)}\bra{q(t)}$. Its expectation value at $t$ reads:
\begin{equation}
\bar{Q}(t) = \text{Tr}\left(\hat{Q}(t)\rho(t)\right)
\end{equation}
Here, we can choose, in particular, $\rho(t)=\rho_c(t)$ or 
$\rho(t)=\rho_{ic}(t)$. Since we are interested in differences between the decay of an UDW detector prepared in a coherent versus an incoherent superposition of velocities, let us consider the difference between the predictions for $\hat{Q}(t)$ for the two density operators
\begin{equation}
    \Delta \bar{Q}(t):= \text{Tr}\left(\hat{Q}(t)\Delta\rho(t)\right),
\end{equation}
where we defined:
\begin{eqnarray}
    \Delta \rho(t)&:=& \rho_c(t)-\rho_{ic}(t)\\
    &=& \frac{1}{2}\sum_{i\neq j} \left(\ket{\Psi_i(0)}\bra{\Psi_j(0)} 
    +\lambda \ket{\psi_i(t)}\bra{\Psi_j(0)} +\lambda \ket{\Psi_i(0)}\bra{\psi_j(t)} \right)
\end{eqnarray}
Noting that all $\ket{\Psi_i(0)}$ and $\ket{\psi_j (t)}$ are mutually orthogonal for $i,j=1,2$, we obtain: 
\begin{eqnarray}
\Delta \bar{Q}(t) &=& \bra{q(t)}\Delta\rho(t)\ket{q(t)}\\ 
&=& \frac{1}{2}\sum_{i\neq j}\left(\alpha_i^*\alpha_j+\lambda s_i(t)\beta^*_i\alpha_j+\lambda s_j(t)\alpha_i^*\beta_j\right)
\end{eqnarray}
To interpret this result, let us first consider the leading terms, $\alpha_i^*\alpha_j$, which possess no $\lambda$-dependence, i.e., which are there even if there is no interaction Hamiltonian or if the interaction time goes to zero, since $s_i(t)\rightarrow 0$ at $t\rightarrow 0$. These terms are present because even at the initial time, when the excited UDW detector and the field have not yet interacted, it is possible to find observables that can tell the difference between the UDW detector being prepared in a coherent or incoherent superposition of velocities. For example, this is the case for any observable $\hat{Q}(t)$ defined using coefficients for which $\alpha_1\neq 0$ and $\alpha_2\neq 0$. Physically, these observables $\hat{Q}(t)$ are binary observables which measure whether the state of motion of the UDW detector is a particular coherent superposition of the two velocities. 
If instead only one of the coefficients $\alpha_i$ is nonzero, then $\hat{Q}$ measures the UDW detector in a particular velocity. This measurement would not be sensitive to whether the UDW detector was prepared in a coherent or incoherent superposition of the two velocities. 

Secondly, there are the terms $\lambda s_i(t)\beta^*_i\alpha_j+\lambda s_j(t)\alpha_i^*\beta_j$ which are proportional to $\lambda$ and that, therefore, arise if the interaction is switched on, i.e., if the UDW detector can decay. The presence of these terms shows that the decay of an excited UDW detector, already to linear order in $\lambda$, does proceed in a way that is sensitive to whether the UDW detector was prepared in a coherent or incoherent superposition of velocities. An observable $\hat{Q}$ can tell the difference if at least one $\alpha_i$ coefficient and one $\beta_j$ coefficient are nonzero, with $i\neq j$. 
The fact that at least one $\alpha_i$ coefficient has to be nonzero is curious because it means that for the observable $\hat{Q}$ to be sensitive to coherent versus incoherent preparation, $\hat{Q}$ cannot be chosen to be such as to merely test whether the UDW detector has decayed yet. Instead, $\hat{Q}$ must be chosen to test for the system to be in a state that is a coherent linear superposition of decayed and still excited.  

To summarize, we have shown that the evolution of the probabilities of decay-related observables can depend on whether the decaying particle was prepared in a coherent or incoherent superposition of momenta. In particular, we showed that this is the case already to first order in perturbation theory when choosing observables whose outcome leaves quantum uncertainty about the total momentum and about whether the atom has or has not decayed.  
Such observables can also be found in full quantum electrodynamics and it should be very interesting to explore these for possible experiments. 

For example, one such observable for a decaying atom could be a location measurement of the atom, such as the measurement of whether, at a time $t$, the decaying atom has passed through a slit. This probability depends on the atom’s initial momentum and, due to a possible recoil, also on whether or not it decayed. At the same time, as required, the outcome of such a position measurement leaves quantum uncertainty regarding the total momentum, the energy and therefore also the decay status. An example for a suitable observable of the field would be a measurement, at time $t$, of the electric field in a region near the atom, using, for example, a quantum homodyne detector. The probabilities of the outcomes of this measurement depend on the total momentum and the decay status (as an emitted photon impacts the field) but, as required, it also leaves quantum uncertainty regarding the decay status and regarding the total momentum. 

We remark here that a related phenomenon of first order perturbation theory in the light-matter interaction has appeared before: the work \cite{firstorderisgood} considered two (then still first quantized) UDW detectors, one to emit a `photon' and one to absorb it. One obtains a quantum channel that maps the initial density matrix of the first UDW detector into the density matrix of the second (after tracing out the first UDW detector and the field). For essentially the same mathematical reasons as above, it was found that the communication channel's leading order in $\lambda$ is obtained by encoding ones and zeros not as either the excited or the ground state of the sending UDW detector but as coherent linear combinations. Else, the leading order terms would be higher order in $\lambda$. This means that the emission of a real photon by an UDW detector is preceded (in the sense of being at lower order in perturbation theory) by the UDW detector and field system evolving into a state which is a linear combination of emission and nonemission. The phenomenon may be viewed as communication through virtual rather than real photons.

\section{Quantum reference frame transformations for field operators}
\label{sec4}
As we discussed in the Introduction, it is often useful to transform into the rest frames of moving systems. For example, one may want to calculate the decay  of an UDW detector in its rest frame and then transform predictions back into the laboratory frame. This would appear to be obstructed when the UDW detector is prepared in a coherent superposition of velocities. Here, we use the formalism of QRFs to build a transformation to the rest frame of an UDW detector even if it is in a coherent superposition of velocities. We then use our generalization of the QRF transformation to field operators in second quantization to describe the process of spontaneous emission in both the rest frame of the UDW detector and in the laboratory frame.

\subsection{The formalism of quantum reference frames}

We now review the essential elements of the formalism introduced in Ref.~\cite{Giacomini:2017zju}, which allows one to build a transformation to a QRF, i.e., a reference frame associated to a quantum system. This formalism is fully relational, meaning that the description of a physical system is given in terms of relative variables to the chosen QRF. A consequence of this fact is that the QRF does not describe itself, but only the relative degrees of freedom between external systems and the QRF. For instance, let us consider three particles $A$, $B$, and $C$. From the perspective of $C$, one describes the degrees of freedom of $A$ and $B$ relative to $C$. From the perspective of $A$, one describes the relative degrees of freedom of $B$ and $C$ relative to $A$. The map between the two QRFs is encoded in a unitary transformation $\hat{S}$, which maps the Hilbert space of $A$ and $B$ relative to $C$, $\mathcal{H}_A^{(C)}\otimes \mathcal{H}_B^{(C)}$ into the Hilbert space of $B$ and $C$ relative to $A$, $\mathcal{H}_B^{(A)}\otimes \mathcal{H}_C^{(A)}$. The quantum state is then transformed unitarily as $\hat{\rho}_{BC}^{(A)} = \hat{S} \hat{\rho}_{AB}^{(C)} \hat{S}^\dagger$, where $\hat{\rho}_{XY}^{(Z)}$, for $X, Y, Z \in \{A, B, C\}$, is the quantum state of $X$ and $Y$ from the perspective of $Z$. A result of Ref.~\cite{Giacomini:2017zju} that is relevant to this work is the realisation that entanglement and superposition are properties that depend on the QRF chosen. In addition, it is possible to build a QRF transformation which is an \emph{extended symmetry} of the Hamiltonian. This means that, if in the initial QRF, say $C$, the quantum systems $A$ and $B$ evolve according to the Hamiltonian 
\begin{equation}
    \hat{H}_{AB}^{(C)} = \hat{H}\left( \lbrace \hat{x}_i, \hat{p}_i, m_i\rbrace_{i=A,B} \right),
\end{equation}
a QRF transformation $\hat{S}$ from $C$ to $A$ corresponding to an extended symmetry transformation maps the initial Hamiltonian to a Hamiltonian of systems $B$ and $C$ such that (see Ref.~\cite{Giacomini:2017zju} for details)
\begin{equation}
    \hat{H}_{BC}^{(A)} = \hat{S} \hat{H}_{AB}^{(C)} \hat{S}^\dagger + i \frac{d \hat{S}}{dt}\hat{S}^\dagger = \hat{H}\left( \lbrace \hat{x}_j, \hat{p}_j, m_j\rbrace_{j=B, C} \right),
\end{equation}
where the operator $\hat{H}$ has the same functional form in both QRFs, but with all labels $A$ and $C$ swapped. For instance, the free-particle Hamiltonian from the perspective of $C$, $ \hat{H}_{AB}^{(C)} = \frac{\hat{p}_A^2}{2m_A} + \frac{\hat{p}_B^2}{2m_B}$ is mapped, via an extended symmetry transformation, into $ \hat{H}_{BC}^{(A)} = \frac{\hat{p}_B^2}{2m_B} + \frac{\hat{p}_C^2}{2m_C}$. Such transformations can be, e.g., \emph{superpositions of spatial translations} or \emph{superpositions of Galilean boosts}.

The probabilities associated to detecting an outcome $b^*$ in the QRF of $C$, which are obtained as $p(b^*) = \text{Tr} \left[\hat{\rho}_{AB}^{(C)} \hat{O}^{(C)}_B(b^*) \right]$, with $\hat{O}^{(C)}_B(b^*)$ being the projector on the outcome $b^*$, are conserved under change of the QRF. However, the QRF transformation changes the partition of the total Hilbert space, and hence one obtains that, in a different QRF, the projection in general acts not only on the Hilbert space of system $B$, but also on the Hilbert space of $C$. The probability is then $p(b^*) = \text{Tr} \left[\hat{\rho}_{BC}^{(A)} \hat{O}^{(A)}_{BC}(b^*) \right]$, where from the perspective of $A$ the projector also acts on $C$ and is related to the one in the QRF of $C$ via $\hat{O}^{(C)}_B(b^*) = \hat{S} \hat{O}^{(C)}_B(b^*) \hat{S}^\dagger$.

A generalisation of this formalism to a special-relativistic particle was achieved in Refs.~\cite{giacomini2019relativistic, streiter2020relativistic}. These works exploit the possibility of transforming to the rest frame of a special-relativistic quantum particle moving in a superposition of momenta via a \emph{superposition of Lorentz boosts}. In particular, such a transformation was used to find a relativistic Stern-Gerlach experiment~\cite{giacomini2019relativistic}, and hence an operational definition of spin at relativistic regimes, and to show that the CHSH Bell inequality can be maximally violated when the particles travel at relatistic velocities, which can also be in a superposition state~\cite{streiter2020relativistic}. In addition, such a violation is independent of the QRF chosen.

\subsection{Introduction of an ancillary system as a quantum reference frame.} 

We now apply the model of the previous section to the study of a spontaneous emission process. In particular, we consider the following situation: in the reference frame of a laboratory $L$ in which the experiment is performed, the atom constituting the Unruh-DeWitt detector has a state which is in a superposition of velocities. In that reference frame, the time when the spontaneous emission happens can be unambiguously identified. However, in an arbitrary laboratory frame, which assigns to the atom a state whose state of motion is not classical in momentum space, but is described as a coherent superposition of velocities, there is no single Lorentz transformation one can apply to the state and to the time of the emission to deduce what an observer at rest in the laboratory frame would measure. It is then non-trivial to describe this situation in the laboratory frame. 

We here answer this question by using tools from QRFs. In particular, we introduce an ancillary degree of freedom $A$ which is needed to derive predictions on the spontaneous emission in the rest frame of the atom. The state of this ancillary system is engineered so that it is perfectly correlated in momentum space with the centre of mass of the atom, in a way that the velocity of the centre of mass of the atom as seen from the ancilla is almost zero. Mathematically, this means that, in the rest frame of the ancilla, the quantum state of D is sharply localised in momentum space
\begin{equation}
     \ket{\Psi_i}^{(A)} = \int \frac{d^3p}{\sqrt{2\omega_{p,1}^D}} \tilde{\delta}(p) b^\dagger_{p,1}\ket{0}_D \ket{0}_E \ket{\chi}_L,
\end{equation}
where the superscript indicates that the state is described in the rest frame of $A$, $\ket{\chi}_L = \int \frac{d^3 q}{\sqrt{2\omega_q^L}}\chi(q) \ket{\vec{q}}_L$, with $\ket{\vec{q}}_L =  d^\dagger_q \ket{0}_L$, is the state of the laboratory from the perspective of the ancilla, and for probability normalisation purposes we describe $\tilde{\delta}(p)$ as a coherent state centred in zero, such that the expectation values of the momentum operator and of the square power of the momentum operator of the atom are approximately zero, i.e., $\langle \hat{p}_D^{(A)} \rangle \approx 0$ and $\langle (\hat{p}_D^{(A)})^2 \rangle \approx 0$.

The full free Hamiltonian also contains the free Hamiltonian of the laboratory from the perspective of the ancilla, and the interaction Hamiltonian is the same as in Eq.~\eqref{HamInt_Sch_UDW}, because the laboratory does not interact with the other systems:

\begin{align}
    \hat{H}_{free}^{(A)} =& \sum_{a=0,1} \int d^3 p \,\omega_{p,a}^D \hat{b}_{p,a}^\dagger \hat{b}_{p,a} + \int d^3 k \,  |k|\, \hat{a}_{\vec{k}}^{\dagger} \hat{a}_{\vec{k}}^{\vphantom \dagger}+ \int d^3 p' \,\omega_{p'}^L \hat{d}_{p'}^\dagger \hat{d}_{p'},\\
   \hat{H}_{int}^{(A)} =& \lambda \int d^3 x \psi^\dagger(x) \hat{\sigma}_1 \psi(x) \Phi (x), \qquad \hat{\sigma}_1 = \begin{pmatrix} 0 & 1\\ 1 & 0 \end{pmatrix}.
\end{align}
Notice that we here start with the full interaction Hamiltonian rather than its rotating wave approximation. When the rest frame of the ancilla coincides with the rest frame of the particle, we approximate $\omega_{p,a}^D \approx m_D + \frac{p_D^{2}}{2m_D}+ E_a + \cdots$. Apart from adding the state of the laboratory from the perspective of the ancilla, nothing else changes in the analysis we carried out in the previous sections, because the laboratory does not interact with the other systems, and only evolves through the free Hamiltonian.

\subsection{Quantum reference frame transformations in second quantization}

We now look for an operator which transforms the description of states and operators from the perspective of the ancilla to the description of states and operators from the perspective of the laboratory. In particular, we want to find a transformation corresponding to the ``superposition of Lorentz boosts'' acting on the Hilbert space of the UDW detector, the electromagnetic field, and the laboratory.

In the interaction picture, such a transformation is (see Appendix~\ref{App:IntPicture})
\begin{equation} 
   \hat{S} = \mathcal{P}^{(v)}_{LA} \hat{U}_D(\Lambda_{-\hat{p}_L/m_L})\hat{U}_{E}(\Lambda_{-\hat{p}_L/m_L}),
\end{equation}
where $\hat{U}_i(\Lambda_{-\hat{p}_L/m_L})$, with $i= D, E$, is the unitary representation of a boost controlled by the velocity of L on the particle and on the electromagnetic field respectively, where $[\hat{U}_D , \hat{U}_E]=0$ because they are both diagonal in the momentum representation of the laboratory. In addition, $ \mathcal{P}^{(v)}_{LA}$ is the generalised parity-swap operator whose representation is 
\begin{equation}
    \mathcal{P}^{(v)}_{LA} =  \mathcal{P}_{LA} \exp \left\lbrace i \log \sqrt{\frac{m_A}{m_L}} (\hat{x}_L \hat{p}_L + \hat{p}_L \hat{x}_L)\right\rbrace
\end{equation}
Here, $\mathcal{P}_{LA}$ is the parity-swap operator acting as $\mathcal{P}_{LA} \hat{x}_L \mathcal{P}_{LA}^\dag = - \hat{q}_A$, $\mathcal{P}_{LA} \hat{p}_L \mathcal{P}_{LA}^\dag = - \hat{\pi}_A$. The momentum operator $\hat{p}_L$ ($\hat{\pi}_A$) is the one-particle operator which is diagonal in the basis $\ket{p}_L$ ($\ket{\pi}_A$), and the operator $\hat{x}_L$ ($\hat{q}_A$) is canonically conjugated to it. The action of the generalised parity-swap operator $ \mathcal{P}^{(v)}_{LA}$ additionally rescales the position and momentum operators, i.e., $\mathcal{P}^{(v)}_{LA} \hat{x}_L \mathcal{P}^{(v)\dag}_{LA} = - \frac{m_A}{m_L}\hat{q}_A$, $\mathcal{P}^{(v)}_{LA} \hat{p}_L \mathcal{P}^{(v)\dag}_{LA} = - \frac{m_L}{m_A}\hat{\pi}_A$, so that the velocity of L as seen from A is equal to the opposite of the velocity of A as seen from L (notice that this condition is the same for both low velocities and  velocities which are close to the speed of light). It can be shown that this definition of the generalised parity-swap operator implies that its action on a momentum eigenstate is $ \mathcal{P}^{(v)}_{LA} \ket{p}_L= \left(\frac{m_A}{m_L}\right)^{3/2} \ket{-\frac{m_A}{m_L} p}_A$, with $\ket{p}_L = d^\dagger_p \ket{0}_L$ (and analogously for $A$).

The operator $\hat{U}_{E}$ (or $\hat{U}_{D}$) is a controlled Lorentz boost on the momentum of the laboratory as seen from the rest frame of the ancilla, and acts on the Hilbert space of the electromagnetic field $E$ (or UDW detector $D$). From the transformation under Lorentz boosts of the creation and annihilation operators in QFT (see Appendix~\ref{App:QFTnotation}), we can find the action of the superposition of boosts on the creation and annihilation operators of $E$ for QRFs
\begin{equation}
    \begin{split}
       & \hat{U}_{E}(\Lambda_{-\hat{p}_L/m_L}) \hat{a}_{k}^{(\dagger)} \hat{U}_{E}^\dagger (\Lambda_{-\hat{p}_L/m_L})= \\
        &=\hat{U}_{E} (\Lambda_{-\hat{p}_L/m_L})\left( \hat{a}_k^{(\dagger)} \otimes \int d^3 p \ket{p}_L \bra{p}
        \right)\hat{U}_{E}^\dagger (\Lambda_{-\hat{p}_L/m_L})= \\
       &= \int d^3 p  \sqrt{\frac{\omega_{\Lambda_{-p} k}}{\omega_p}} \hat{a}_{\Lambda_{-p} k}^{(\dagger)} \ket{p}_L \bra{p},
    \end{split}
\end{equation}
where $\int d^3p \ket{p}_L \bra{p}$ is a resolution of the identity in the one-particle sector of the Hilbert space of the laboratory. 

Since the Lorentz boost is a symmetry of spacetime, it leaves the vacuum state of the field invariant. As a consequence, the operator $\hat{U}_{E}$ also leaves the vacuum state of the EM field invariant. To see this, it is sufficient to introduce the resolution of the identity used earlier and to notice that, when applied to the state $\ket{p}_L = d^\dagger_p \ket{0}_L$, the operator $\hat{U}_{E}$ acts as a normal Lorentz boost by a definite velocity. We can then write a general expression for the action of $\hat{U}_{E}$ on the vacuum state and on a one-particle state, i.e.,
\begin{align}
    &\hat{U}_{E} \ket{0}_E = \ket{0}_E;\\
    &\hat{U}_{E} \hat{a}_k^\dagger \ket{0}_E = \sqrt{\frac{\omega_{\Lambda_{-p} k}}{\omega_p}} \hat{a}_{\Lambda_{-p} k}^\dagger \ket{0}_E.
\end{align}

The superposition of Lorentz boosts $\hat{U}_D(\Lambda_{-\hat{p}_L/m_L})$ works analogously for the UDW detector D. Here, the operator $\hat{U}_D(\Lambda_{-\hat{p}_L/m_L})$ acts on the full Hilbert space of the particle, and hence on both the external degrees of freedom D and the internal degrees of freedom $I$. In order to transform the state, the two components corresponding to the ground and excited state of the internal degrees of freedom transform independently:
\begin{equation}
    \begin{split}
        &\hat{U}_{D}(\Lambda_{-\hat{p}_L/m_L}) \hat{b}_{p,a}^{(\dagger)} \hat{U}_{D}^\dagger (\Lambda_{-\hat{p}_L/m_L})= \\
       &= \int d^3 \pi  \sqrt{\frac{\omega^D_{\Lambda_{-\pi} p, a}}{\omega^D_{p,a}}} \hat{b}_{\Lambda_{-\pi} p,a}^{(\dagger)} \ket{\pi}_L \bra{\pi}, \qquad \text{for} \qquad a=0,1.
    \end{split}
\end{equation}
In particular, the action of the operator $\hat{U}_{D}(\Lambda_{-\hat{p}_L/m_L})$ on the vacuum state of the particle is $\hat{U}_{D}(\Lambda_{-\hat{p}_L/m_L}) \ket{0}_D = \ket{0}_D$.

\subsection{Transformation of the interaction Hamiltonian}

The transformation of the interaction Hamiltonian to the laboratory frame in the interaction picture reads, remebering that $\psi(x)$ and $\Phi (x)$ are scalar fields
\begin{equation} \label{eq:IntHamiltL}
   \begin{split}
        \hat{H}_{int}^{(L)} &= \hat{\gamma}_A^{-1} \hat{S} \hat{H}_{int}^{(A)} \hat{S}^\dagger=\\
        &=\lambda \int d^3 x d^3\pi \left(\gamma_\pi^A\right)^{-1}\,\psi^\dagger\left( \Lambda_{\pi/m_A} x\right)\hat{\sigma}_1 \psi\left( \Lambda_{\pi/m_A} x\right)  \Phi \left( \Lambda_{\pi/m_A} x\right) \ket{\pi}_A \bra{\pi}=\\
        &= \lambda \int d^3 x' d^3\pi \psi^\dagger\left( x' \right) \hat{\sigma}_1  \psi\left(  x' \right) \Phi \left( x' \right) \ket{\pi}_A \bra{\pi}=\\
         &= \lambda \int d^3 x' \psi^\dagger\left( x' \right)\hat{\sigma}_1  \psi\left(  x' \right) \Phi \left( x' \right),
   \end{split}
\end{equation}
where $\hat{\gamma}_A = \sqrt{1+ \frac{\hat{p}^2_A}{m_A^2}}$, $\hat{p}_A$ is the momentum operator acting on the one-particle sector of the ancilla Hilbert space, and we have used the fact that $d^3 x = \gamma_\pi^A d^3 x'$, with $\gamma_\pi^A$ being the eigenvalue of the operator $\hat{\gamma}_A$ when evaluated on a momentum eigenstate $\ket{\pi}_A = \hat{d}_\pi^\dagger \ket{0}_A$. From the previous result, we see that the functional form of the interaction Hamiltonian in the laboratory frame is the same as in the rest frame of the ancilla. 

We can now recover the resonant interaction Hamiltonian in the laboratory frame. The procedure is covariant, as it corresponds to the conservation of the $4$-momentum and to dropping the counter-rotating terms. We thus have the four constraints $p' = p+k$, where the time component of the $4$-vector is $\omega_{p', 1}^D = \omega_{p,0}^D + |k| $. With this, the interaction Hamiltonian is, in the laboratory frame:
\begin{equation}
    \hat{H}_{int}^{(L, res)} =  \frac{\lambda}{(2\pi)^{3/2}} \int \frac{d^3 k}{\sqrt{2|k|}} \frac{d^3 p}{\sqrt{2\omega_{p,0}^D}} \frac{1}{\sqrt{2\omega_{p+k, 1}^D}} \, \left[ e^{i (\omega_{p,0} +|k| - \omega_{p+k,1})t}\, b^\dagger_{p,0} a^\dagger_k b_{p+k,1}  + h.c.\right].
\end{equation}

\subsection{Quantum reference frame transformation of the states}

The quantum state is transformed to the laboratory frame by applying the operator $\hat{S}$ in the interaction picture to the state $\ket{\Psi}^{(A)}$ of Eq.~\eqref{eq:FullStateinA}. We obtain
\begin{equation} \label{eq:FullStateinL}
   \ket{\Psi}^{(L)}= \hat{S} \ket{\Psi}^{(A)} = \ket{\Psi_i}^{(L)} -i \lambda \ket{\Psi_f}^{(L)},
\end{equation}
where $\ket{\Psi_i}^{(L)} = \hat{S} \ket{\Psi_i}^{(A)}$ and $\ket{\Psi_f}^{(L)} = \hat{S} \ket{\Psi_f}^{(A)}$ and
\begin{align}
     \ket{\Psi_i}^{(L)} = &\frac{m_L}{m_A} \int \frac{d^3 p}{\sqrt{2\omega_{p,1}^D}}\frac{d^3 \pi}{\sqrt{2\omega_\pi^A}}\tilde{\delta}\left(\Lambda_{\pi/m_A}^{-1} p\right)\chi\left(-\frac{m_L}{m_A} \pi\right) b^\dagger_{p,1} \ket{0}_D d^\dagger_\pi \ket{0}_A \ket{0}_E \\
     \ket{\Psi_f}^{(L)} = &\frac{1}{(2\pi)^{3/2}}\frac{m_L}{m_A} \sum_a \int \frac{d^3 p}{\sqrt{2\omega_{p,a}^D}}\frac{d^3 \pi}{\sqrt{2\omega_\pi^A}}\frac{d^3 k}{\sqrt{2|k|}} \frac{1}{2\omega_{p+k,1}^D} \tilde{\delta}\left(\Lambda_{\pi/m_A}^{-1} (p+k) \right) \times\\ 
     & \chi\left(-\frac{m_L}{m_A}\pi\right) b^\dagger_{p,a} \ket{0}_D d^\dagger_\pi \ket{0}_A a^\dagger_k \ket{0}_E , \nonumber
\end{align}
where $x'\,^0 = t-t_i$ and we have used that the spacetime integral is an invariant quantity. The relation between proper time and laboratory time is, for each specific momentum $\pi$ of the ancilla in the laboratory frame
\begin{equation}
   c t = \gamma_\pi (c\tau - \beta_\pi x) ; \qquad c t_i = \gamma_\pi (c \tau_i - \beta_\pi x_i).
\end{equation}

Here, a quantum state evolving to a time $\tau_f$ in the rest frame of the ancilla corresponds to a quantum state evolving to a quantum superposition of laboratory times, namely:
\begin{align}
       \ket{\Psi}^{(L)} = \hat{S} \ket{\Psi}^{(A)} &=  \hat{S} \left[\ket{\Psi_i}^{(A)} -i\int_{\tau_i}^{\tau_f} d \tau \hat{H}_{int}^{(A)}(\tau-\tau_i)\ket{\Psi_i}^{(A)} \right]\\
        &=\ket{\Psi_i}^{(L)} -i \int_{\hat{\gamma}_A \tau_i}^{\hat{\gamma}_A \tau_f} d t \hat{\gamma}_A^{-1} \hat{S} \hat{H}_{int}^{(A)}(\tau-\tau_i)\hat{S}^\dagger \ket{\Psi_i}^{(L)}\\
         &=\ket{\Psi_i}^{(L)} -i \int_{\hat{\gamma}_A \tau_i}^{\hat{\gamma}_A \tau_f} d t \hat{H}_{int}^{(L)}(t-t_i)\ket{\Psi_i}^{(L)}. \label{xyz}
\end{align}
Here, let us recall that the Lorentz factor $\gamma_{\hat{\pi}_A}$ is an operator on the Hilbert space of the ancilla and it acts here on the part of $\ket{\Psi_i}^{(L)}$ which is in the Hilbert space of the ancilla. 
Therefore, $\gamma_{\hat{\pi}_A}$ becomes effectively number valued when the state of the ancilla is expanded in the momentum basis. If the state of the ancilla is chosen to be a generic coherent or incoherent linear combination of momentum eigenstates, Eq.~\eqref{xyz} yields states at the range of times $t$ that are the image of the point in time $\tau$.  If the ancilla is instead chosen to be in a momentum eigenstate, we recover a definite Lorentz transformation, where the state in the laboratory frame is at a single time. 

\section{Application: Rate operator}
\label{sec5}

Let us consider an UDW detector in an arbitrary frame $R$ interacting with the electromagnetic field. We now wish to describe the spontaneous emission of a photon from an UDW detector moving in a superposition of relativistic velocities. Usually, we use the spontaneous emission rate to characterize emission processes. We see in this Section that, when the particle emitting a photon is in a superposition of relativistic velocities, we need to introduce a rate operator. The transformation properties of such a rate operator between two different QRFs are not trivial and yield some interesting consequences. We work in the interaction picture.

We start by deriving the rate operator. In an initial reference frame, say $R$, we define a state at time $t$ of the UDW detector, the electromagnetic field, and the ancilla as $\hat{\rho}^{(R)}$, and a measurement (which can be a projective measurement or a Positive Operator-Valued Measure--POVM) as $\hat{\Pi}^{(R)}$. Probabilities are defined as
\begin{equation}
    P = \text{Tr}\left[\hat{\rho}^{(R)} \hat{\Pi}^{(R)}\right].
\end{equation}
For instance, we could take $\hat{\Pi}^{(R)}$ to be a projector on the one-photon Hilbert space. With this, we calculate the probability of the spontaneous emission having occurred. The transition rate $\mathcal{R}^{(R)}$ calculated in the initial frame $R$ is the time derivative of the probability, namely
\begin{equation}
    \mathcal{R}^{(R)} = \frac{d}{dt}\text{Tr}\left[\hat{\rho}^{(R)} \hat{\Pi}^{(R)}\right].
\end{equation}
We can rewrite the previous expression as
\begin{align}
    \mathcal{R}^{(R)} &= \text{Tr}\left[\frac{d}{dt}\left(\hat{\rho}^{(R)}\right) \hat{\Pi}^{(R)} + \hat{\rho}^{(R)}\frac{d}{dt}\left( \hat{\Pi}^{(R)}\right)\right]=\\
    &= \text{Tr}\left[  \hat{\rho}^{(R)}\frac{d \hat{\Pi}^{(R)}}{dt} - i \left[ \hat{H}_{int}^{(R)}, \hat{\rho}^{(R)} \right] \hat{\Pi}^{(R)}\right]=\\
      &= \text{Tr}\left[ \left(\frac{d \hat{\Pi}^{(R)}}{dt}  + i \left[ \hat{H}_{int}^{(R)}, \hat{\Pi}^{(R)} \right] \right) \hat{\rho}^{(R)} \right],
\end{align}
where we have used the cyclicity of the trace. We hence define the rate operator in an arbitrary frame $R$ as
\begin{equation}
    \hat{\mathcal{R}}^{(R)} = \frac{d \hat{\Pi}^{(R)}}{dt}  + i \left[ \hat{H}_{int}^{(R)}, \hat{\Pi}^{(R)} \right].
\end{equation}

In the situation we have considered in this work, we consider the rate operators in the laboratory frame and in the ancilla frame, i.e., $R \in \{A, L\}$. We now want to show the relation between the two rate operators under a QRF transformation. Knowing that states and observables transform under a QRF transformation $\hat{S}$ from $A$ to $L$ as $\hat{\rho}^{(L)} = \hat{S} \hat{\rho}^{(A)} \hat{S}^\dagger$ and $\hat{\Pi}^{(L)} = \hat{S} \hat{\Pi}^{(A)} \hat{S}^\dagger$ respectively, we find
\begin{align}
    \mathcal{R}^{(A)}  &=  \text{Tr}\left[\hat{\mathcal{R}}^{(A)} \hat{\rho}^{(A)}\right]= \text{Tr}\left[ \left(\frac{d \hat{\Pi}^{(A)}}{d\tau_A}  + i \left[ \hat{H}_{int}^{(A)}, \hat{\Pi}^{(A)} \right] \right) \hat{\rho}^{(A)} \right]=\\
    &= \text{Tr}\left[ \hat{S} \left(\frac{d \hat{\Pi}^{(A)}}{d\tau_A}  + i \left[ \hat{H}_{int}^{(A)}, \hat{\Pi}^{(A)} \right] \hat{S}^\dagger \right) \hat{\rho}^{(L)} \right]=\\
     &= \text{Tr}\left[ \left( \hat{S}\frac{d \hat{\Pi}^{(A)}}{d\tau_A}\hat{S}^\dagger  + i \left[ \hat{S}\hat{H}_{int}^{(A)} \hat{S}^\dagger, \hat{\Pi}^{(L)} \right] \hat{S}^\dagger \right) \hat{\rho}^{(L)} \right],
\end{align}
where $\tau_A$ is the proper time in the rest frame of the ancilla and we have used $\hat{S} \hat{S}^\dagger = \mathbb{1}$ and the cyclicity of the trace. By using the identity
\begin{equation}
    \hat{\gamma}_A \frac{d (\hat{S} \hat{\Pi}^{(A)}\hat{S}^\dagger)}{dt_L} =\hat{S}\frac{d \hat{\Pi}^{(A)}}{d\tau_A}\hat{S}^\dagger + \left[  \hat{\gamma}_A \frac{d \hat{S}}{dt_L} \hat{S}^\dagger, \hat{\Pi}^{(L)}\right],
\end{equation}
where the time in the ancilla frame $\tau_A$ and in the laboratory frame $t_L$ are related by $\Delta \tau_A = \hat{\gamma}_A^{-1} \Delta t_L$, we can write the expression of the rate $\mathcal{R}^{(A)}$ as
\begin{align}
    \text{Tr}\left[ \mathcal{\hat{R}}^{(A)}\hat{\rho}^{(A)} \right] &= \text{Tr}\left[ \left( \hat{\gamma}_A\frac{d \hat{\Pi}^{(L)}}{dt_L}  + i \left[ \hat{\gamma}_A \hat{H}_{int}^{(L)}, \hat{\Pi}^{(L)} \right]- \left[\hat{\gamma}_A \frac{d\hat{S}}{dt_L} \hat{S}^\dagger, \hat{\Pi}^{(L)} \right] \right) \hat{\rho}^{(L)} \right]=\\
    &= \text{Tr}\left[ \left( \hat{\gamma}_A \hat{\mathcal{R}}^{(L)}  + i \left[ \hat{\gamma}_A, \hat{\Pi}^{(L)} \right] \hat{H}_{int}^{(L)}- \left[\hat{\gamma}_A \frac{d\hat{S}}{dt_L} \hat{S}^\dagger, \hat{\Pi}^{(L)} \right] \right) \hat{\rho}^{(L)} \right].
\end{align}
Conversely, we also have
\begin{equation}\label{eq:RateOpQRFTrasf}
    \text{Tr}\left[ \mathcal{\hat{R}}^{(L)}\hat{\rho}^{(L)} \right] = \text{Tr}\left[ \left( \hat{\gamma}_L \hat{\mathcal{R}}^{(A)}  + i  \left[ \hat{\gamma}_L, \hat{\Pi}^{(A)} \right] \hat{H}_{int}^{(A)}- \left[\hat{\gamma}_L \frac{d\hat{S}^\dagger}{d\tau_A} \hat{S}, \hat{\Pi}^{(A)} \right] \right) \hat{\rho}^{(A)} \right],
\end{equation}
where $\hat{S}^\dagger$ is the inverse transformation to the one considered so far, performing a superposition of Lorentz boosts from the QRF of $L$ to the QRF of $A$. We see that there are two additional terms to those we expect in a standard Lorentz boost. One vanishes if the QRF transformation is time-independent, which is verified for the quantum superposition of boosts in the interaction picture. More interestingly, the other term vanishes if $\left[ \hat{\gamma}_L, \hat{\Pi}^{(A)} \right] =0$, e.g., if the measurement operator is diagonal in momentum basis of the ancilla Hilbert space (whose state is perfectly correlated in momentum basis with the UDW detector). This condition is always satisfied for standard reference frames transformation, but the projector can be chosen not to vanish for a QRF transformation.

In summary, we recover the standard relation between rates in different Lorentz frames when the velocity of the particle is well-defined, or when the operator $\hat{\Pi}^{(A)}$ has no coherence terms in momentum basis in the Hilbert space of the laboratory $L$ (i.e., it measures a statistical mixture of velocities). Otherwise, the transformation of the rate operators acquires additional terms, which are purely due to the coherent superposition of velocities of the QRF.

This fact is easily illustrated with two examples of measurements in the QRF of $L$ and $A$. Let us consider a POVM on the Hilbert space of the UDW detector and the photon $\hat{M}_{DE}$. With this, we build two different measurements in the laboratory frame
\begin{align}
    & \hat{\Pi}_1^{(L)} = \int \frac{d^3 p}{2\omega_p^A} f_1(p)  \hat{U}_D \left(\Lambda_{p/m_A}\right)\hat{U}_E \left(\Lambda_{p/m_A}\right) \hat{M}_{DE} \hat{U^\dagger}_E \left(\Lambda_{p/m_A}\right)\hat{U}^\dagger_D \left(\Lambda_{p/m_A}\right) \ket{p}_A \bra{p},\\
     & \hat{\Pi}_2^{(L)} = \int \frac{d^3 p}{\sqrt{2\omega_p^A}}\frac{d^3 p'}{2\omega_{p'}^A} f_2(p, p') \hat{U}_D \left(\Lambda_{p/m_A}\right)\hat{U}_E \left(\Lambda_{p/m_A}\right) \hat{M}_{DE} \hat{U^\dagger}_E \left(\Lambda_{p'/m_A}\right)\hat{U}^\dagger_D \left(\Lambda_{p'/m_A}\right) \ket{p}_A \bra{p'},
\end{align}
where $f_1(p) = \frac{m_A}{m_L}\left|\sigma_1 \left(- \frac{m_L}{m_A}p\right)\right|^2$ and $f_2(p, p') = \frac{m_A}{m_L} \sigma_2 \left(- \frac{m_L}{m_A}p\right)\sigma_2^* \left(- \frac{m_L}{m_A}p'\right)$ and $\sigma_i$, $i=1,2$ are arbitrary wavefunctions of quantum states. The transformation of these two observables to the ancilla frame gives
\begin{align}
    & \hat{\Pi}_1^{(A)} = \int \frac{d^3 \pi}{2\omega_\pi^L} \left|\sigma_1 \left(\pi\right)\right|^2 \ket{p}_L \bra{p} \otimes \hat{M}_{DE},\\
     & \hat{\Pi}_2^{(A)} = \int \frac{d^3 \pi}{\sqrt{2\omega_\pi^L}}\frac{d^3 \pi'}{2\omega_{\pi'}^L} \sigma_2 \left(\pi\right)\sigma_2^* \left(\pi'\right)  \ket{\pi}_L \bra{\pi'} \otimes \hat{M}_{DE}.
\end{align}
It is easy to see, by inserting these two observables in Eq.~\eqref{eq:RateOpQRFTrasf}, that $\left[ \hat{\gamma}_L, \hat{\Pi}_1^{(A)} \right] =0$ and $\left[ \hat{\gamma}_L, \hat{\Pi}_2^{(A)} \right] \neq0$. In the laboratory frame, $\hat{\Pi}_1^{(L)}$ has no coherence terms in the momentum basis of the ancilla, but only a statistical mixture of the different momenta. Instead, $\hat{\Pi}_2^{(L)}$ contains coherent superpositions of different momenta.

\section{Conclusions and Outlook}

We introduced second-quantized Unruh-DeWitt detectors and we generalized the formalism of quantum reference frame transformations to allow the second-quantized UDW detectors to serve to define quantum reference frames.   

We illustrated the use of second-quantized UDW detectors with the example of the decay of excited UDW detectors that are in coherent superposition of relativistically differing velocities. We found that already to first order in the coupling constant $\lambda$ there are decay-related observables that are sensitive to the difference between the incoherent and coherent case. The generalization to higher orders in $\lambda$ should be straightforward. 

The new formalism of second-quantized Unruh-DeWitt detectors and the correspondingly generalized formalism of quantum reference frames should be directly applicable 
to numerous scenarios in which previously first-quantized UDW detectors have been used. This should be useful because it allows one, in particular, to include the effects of UDW detectors being in a coherent superposition of relativistically differing velocities. 

In this context, it is curious that, in the literature, slight but seemingly persistent discrepancies have been reported between theoretically predicted and experimentally observed aspects of the decay of unstable particles, in particular, in positronium and in neutron decay, see, e.g.,  \cite{zatorski2008m, gurung2020precision, space2021lawrence, measurement2021wilson, improved2021gonzalez, Giacosa:2018dzm}. It should be interesting to explore if these discrepancies could be related to assumptions about incoherent versus coherent superpositions of the velocities of the decaying particles. In the context of propagation with a superposition of differing velocities, our approach may also be applicable to the phenomenon of neutrino oscillations. For related work on the use of UDW detectors for the description of neutrino oscillations see \cite{bruno}. 

The formalism of second-quantized Unruh-DeWitt detectors that we introduced here should also be able to describe detectors that are in a superposition of accelerations, see~\cite{Barbado:2020snx, Foo:2020xqn}, namely by exposing them to correspondingly accelerating background fields. For the nonrelativistic case, see~\cite{sudhir2021unruh}.

Of particular interest should be applications of the second-quantized UDW detectors, for example, to entanglement harvesting, quantum communication through quantum fields and to the recently-found effects of acceleration-induced transparency and stimulation of the Unruh effect \cite{soda2021accelerationinduced}. For the latter effects, not only the rotating wave terms of 
the simplified Hamiltonian $H_{int}^{(res)}$ in Eq.~\eqref{Hint} are needed. Crucial in this case are also the counter-rotating terms of the full interaction Hamiltonian of Eq.~\eqref{HamInt_Sch_UDW}. 

In this context, let us recall that the full Hamiltonian Eq.~\eqref{HamInt_Sch_UDW} now also allows the description of the resonant creation and annihilation of UDW detectors, when the energy of the emitted or absorbed quanta are in the range of the rest masses of UDW detectors. 
In addition, the full Hamiltonian Eq.~\eqref{HamInt_Sch_UDW} also contains non-resonant terms describing UDW detector creation (and annihilation), analogous to the well known creation of particles by accelerated UDW detectors in the Unruh effect. 
This indicates that the recent findings in Ref.~\cite{soda2021accelerationinduced} should apply here also for the creation and annihilation of UDW detectors themselves: resonant, i.e., rotating wave terms can be suppressed and non-resonant, i.e., counter-rotating terms can be stimulated by suitable external driving agents such as classical background electromagnetic or gravitational fields. Analogous to the acceleration-induced transparency found in  Ref.~\cite{soda2021accelerationinduced}, 
there may exist, for example, a phenomenon of external-agent induced avoidance of UDW detector creation even if incoming particles are sufficiently energetic to provide the required rest mass energy for the creation of new UDW detectors. 

Finally, let us recall that we naturally arrived at results, in particular in Eq.~\eqref{xyz}, which yielded, in the laboratory frame, quantum state information at a range of times, which may be viewed as a form of \emph{quantum superposition of states at varying times}. This poses the highly interesting challenge to determine how one can optimally extract predictions from state information which is given in this form. 

The generalization of the formalism of quantum reference frame transformations to field operators can be applied to quantum fields (such as the electromagnetic field) with an arbitrary number of particles. 
However, for the purpose of defining a quantum reference frame we have, so far, only made use of the one-particle sector of the quantum field theoretic description of the UDW detector system. It will be interesting to explore steps towards a more general formulation of quantum reference frames that are associated to an arbitrary state of a quantum field~\cite{dewitt1967quantum, Kuchar:1990vy, Brown:1994py, Brown:1995fj}. 

\section{Acknowledgements}
The authors thank Maria Papageorgiou, Tales Rick Perche, and Jos{\'e} Polo-G{\'o}mez for very valuable feedback. AK acknowledges support through a Discovery Grant from the National Science and Engineering Research Council (NSERC) of Canada, a Discovery Grant from the Australian Research Council (ARC) and a Google Faculty Research Award.
The authors acknowledge support from Perimeter Institute for Theoretical Physics. Research at Perimeter Institute is supported in part by the Government of Canada through the Department of Innovation, Science and Economic Development and by the Province of Ontario through the Ministry of Colleges and Universities.

\appendix

\section{Summary of normalisation conventions}
\label{App:QFTnotation}

We take here the non-covariant normalisation of QFT, which prescribes the following commutation relations between the creation and annihilation operators
\begin{equation}
    [\hat{a}_k, \hat{a}_{k'}^\dagger]= \delta(k-k'). 
\end{equation}
A free field is expanded in the following way
\begin{equation}
    \Phi (x) = \frac{1}{(2\pi)^{3/2}} \int \frac{d^3 p}{\sqrt{2\omega_p}}\left\lbrace \hat{a}_p e^{-i p \cdot x} + \hat{a}_p^\dagger e^{i p \cdot x} \right\rbrace
\end{equation}
where $\omega_p = \sqrt{\vec{p}^2 +m^2 c^2}= c^{-1}p^0$, $\vec{p}$ is the spatial part of the momentum, and $p \cdot x$ is the scalar product between two four-vectors. The free Hamiltonian is written as
\begin{equation}
    \hat{H}_{free} = \int d^3 p \omega_p \hat{a}_p^\dagger \hat{a}_p.
\end{equation}
When describing the external degree of freedom of the atom and the ancilla, we use the projection of the Hamiltonian on the one-particle sector $\hat{H}_{free} = \int d^3 p \omega_p  \hat{a}_p^\dagger \ket{0}\bra{0} \hat{a}_p$, which is formally obtained by inserting a resolution of the identity in Fock space in the Hamiltonian, i.e., $\mathbb{1} = \ket{0}\bra{0} + \sum_{n=1}^\infty \frac{1}{n!} \int d^3 p_1 \cdots d^3 p_n a^\dagger_{p_1}\cdots a^\dagger_{p_n}\ket{0}\bra{0}a_{p_1}\cdots a_{p_n}$, and neglecting all contributions except for the vacuum state term.

In order for the theory to be covariant under the Poincar{\'e} group, and in particular under a Lorentz boost $\hat{U}(\Lambda_k)$ we need to have
\begin{align}
   & \hat{U}(\Lambda_k) \hat{a}_p^{(\dagger)}\hat{U}^\dagger(\Lambda_k) = \sqrt{\frac{2\omega_{\Lambda p}}{2\omega_p}}\hat{a}_{\Lambda p}^{(\dagger)},\\
   & \ket{\psi} = \int \frac{d^3 p}{\sqrt{2\omega_p}}\psi(p)\ket{p}, \qquad \text{(one-particle state)}\\
   & \braket{p'|p} = \delta(p'-p), \qquad \text{with} \ket{p}= \hat{a}^\dagger_p \ket{0}\\
   & \mathbb{1} = \int d^3 p \ket{p}\bra{p}, \qquad \text{(identity operator in the one-particle sector)}\\
   & \text{Tr}[\cdots] = \int d^3 p \bra{p} [\cdots] \ket{p}. 
\end{align}

Here, we use the same expressions for the particles and for the electromagnetic field. This amounts to neglecting the polarisation of the photon, and subsequently the Wigner rotation when we apply the boost (notice that for collinear boosts, i.e., in 1+1 dimensions, the Wigner rotation is zero, so we might want to assume that we are in that case).

\section{The QRF transformation in the interaction picture}
\label{App:IntPicture}

In this Appendix, we first look for a transformation in the Schr{\"o}dinger picture representing a ``superposition of Lorentz boosts'' in second quantization, and we then show how to map this transformation to the interaction picture. 

Such a ``superposition of Lorentz boosts'' needs to be an \emph{extended symmetry transformation}, as defined in Ref.~\cite{Giacomini:2017zju}, i.e., it should map the free Hamiltonian of the UDW detector, the electromagnetic field, and the laboratory in the ancilla frame
\begin{equation} \label{eq:FreeHamA}
    \hat{H}_{free}^{(A)} =  \sum_{a=0,1} \int d^3 p\, \omega_{p,a}^D b^\dagger_{p,a} b_{p,a}+   \int d^3 k |k| a^\dagger_k a_k +  c\int d^3 q \omega_q^A d^\dagger_q  d_q.
\end{equation}
to a Hamiltonian of the UDW detector, the electromagnetic field, and the ancilla in the laboratory frame, having the same functional form as the initial one, i.e.,
\begin{equation}
    \hat{H}_{free}^{(L)} =  \sum_{a=0,1} \int d^3 p\, \omega_{p,a}^D b^\dagger_{p,a} b_{p,a}+  \int d^3 k |k| a^\dagger_k a_k +  \frac{m_L}{m_A}\int d^3 q \omega_q^A d^\dagger_q  d_q.
\end{equation}
In the previous expression, the factor $\frac{m_L}{m_A}$ in front of the Hamiltonian of the ancilla is a feature of the rescaling of the momentum operator in the one-particle sector typical of a QRF transformation, but has no consequences for the results contained in this work. In the Schr{\"o}dinger picture, the  transformation of the free Hamiltonian~\eqref{eq:FreeHamA}  from the QRF of the ancilla to the QRF of the laboratory is \cite{Giacomini:2017zju, giacomini2019relativistic}:
\begin{equation}
    \hat{H}_{free}^{(L)} = \hat{\gamma}^{-1}_A \left[ \hat{S}_S \hat{H}_{free}^{(A)}\hat{S}_S^\dagger + i \frac{d \hat{S}_S}{d\tau}\hat{S}_S^\dagger \right].
\end{equation}
The Lorentz factor $\hat{\gamma}_A^{-1}$ comes from the transformation $d\tau = \hat{\gamma}_A^{-1} dt$, where $\tau$ is the proper time in the rest frame of the ancilla and $t$ is the coordinate time (i.e., the proper time in the laboratory frame), and $\hat{\gamma}_A = \sqrt{1+ \frac{\hat{p}_A^2}{m_A^2}}$ is an operator acting on the external degrees of freedom of the ancilla, where $\hat{p}_A$ and $m_A$ are respectively the momentum operator and the mass of the ancilla.

We then find that the transformation corresponding to the ``superposition of Lorentz boosts'' which is also an \emph{extended symmetry transformation} of the free Hamiltonian is, in the Schr{\"o}dinger picture 
\begin{equation} \label{eq:SchrQRFtrans}
    \hat{S}_S = e^{-i \hat{H}_{free}^{(L)}\hat{\gamma}_A \Delta \tau} \mathcal{P}^{(v)}_{LA} \hat{U}_D(\Lambda_{-\hat{p}_L/m_L})\hat{U}_{E}(\Lambda_{-\hat{p}_L/m_L})  e^{i \hat{H}_{free}^{(A)} \Delta \tau},
\end{equation}
where $\hat{U}_i(\Lambda_{-\hat{p}_L/m_L})$, with $i= D, E$, is the unitary representation of a boost controlled by the velocity of L on the particle and on the electromagnetic field respectively, where $[\hat{U}_D , \hat{U}_E]=0$ because they are both diagonal in the momentum representation of the laboratory. The action of the boost operators on the creation and annihilation operators of the field is the one presented in the main text. In addition, $ \mathcal{P}^{(v)}_{LA}$ is the generalised parity-swap operator whose representation is 
\begin{equation}
    \mathcal{P}^{(v)}_{LA} =  \mathcal{P}_{LA} \exp \left\lbrace i \log \sqrt{\frac{m_A}{m_L}} (\hat{x}_L \hat{p}_L + \hat{p}_L \hat{x}_L)\right\rbrace
\end{equation}
Here, $\mathcal{P}_{LA}$ is the parity-swap operator acting as $\mathcal{P}_{LA} \hat{x}_L \mathcal{P}_{LA}^\dag = - \hat{q}_A$, $\mathcal{P}_{LA} \hat{p}_L \mathcal{P}_{LA}^\dag = - \hat{\pi}_A$. The action of the generalised parity-swap operator $ \mathcal{P}^{(v)}_{LA}$ additionally rescales the position and momentum operators, i.e., $\mathcal{P}^{(v)}_{LA} \hat{x}_L \mathcal{P}^{(v)\dag}_{LA} = - \frac{m_A}{m_L}\hat{q}_A$, $\mathcal{P}^{(v)}_{LA} \hat{p}_L \mathcal{P}^{(v)\dag}_{LA} = - \frac{m_L}{m_A}\hat{\pi}_A$, so that the velocity of L as seen from A is equal to the opposite of the velocity of A as seen from L (notice that this condition is the same for both low velocities and  velocities which are close to the speed of light). It can be shown that this definition of the generalised parity-swap operator implies that its action on a momentum eigenstate is $ \mathcal{P}^{(v)}_{LA} \ket{p}_L= \left(\frac{m_A}{m_L}\right)^{3/2} \ket{-\frac{m_A}{m_L} p}_A$.

The Hamiltonian in the rest frame of the ancilla is composed of a free part $\hat{H}^{(A)}_{free}$ and of an interacting part $\hat{H}^{(A)}_{int}$. The total Hamiltonian is $\hat{H}^{(A)}_{tot} = \hat{H}^{(A)}_{free} + \hat{H}^{(A)}_{int}$.  The relation between the Schr{\"o}dinger and the interaction picture is
\begin{align}
    & \hat{\rho}_0^{(A)} = e^{i\hat{H}_{free}^{(A)}\Delta \tau} \hat{\rho}^{(A)}e^{-i\hat{H}_{free}^{(A)}\Delta\tau},\\
    & \hat{H}_{int}^{(A)}(\Delta\tau) = e^{i\hat{H}_{free}^{(A)}\Delta\tau} \hat{H}_{int}^{(A)}e^{-i\hat{H}_{free}^{(A)}\Delta \tau},\\
    & i \frac{d \hat{\rho}_0^{(A)}}{d\tau} = \left[\hat{H}_{int}^{(A)}(\Delta\tau), \hat{\rho}_0^{(A)} \right],
\end{align}
where the subscript $0$ indicates the interaction picture, $\tau$ is the proper time in the rest frame of the ancilla, and $\Delta\tau = \tau_f - \tau_i$. In addition, let's call $\hat{S}_S$ the QRF transformation in the Schr{\"o}dinger picture (for the time being, we won't specify what this transformation is, as the result holds for any unitary transformation). 

Finally, the transformed Hamiltonians in the laboratory frame can be split in the following way
\begin{align}
     &\hat{H}^{(L)}_{free} = \hat{\gamma}_A^{-1} \left(\hat{S}_S \hat{H}^{(A)}_{free} \hat{S}_S^\dagger + i \frac{d \hat{S}_S}{d\tau} \hat{S}_S^\dagger \right) \\
     &\hat{H}^{(L)}_{int} = \hat{\gamma}_A^{-1}\hat{S}_S \hat{H}^{(A)}_{int} \hat{S}_S^\dagger,
\end{align}
where in the most general case (which we do not consider here) the term $\hat{H}^{(L)}_{free}$ does not need to be free of any interaction, but only the relevant one for the spontaneous emission. 

The covariance of the laws of physics implies that we the transformation between the Schr{\"o}dinger and the interaction picture is formally the same in any (quantum) reference frame. Hence, the transformation between the Schr{\"o}dinger and the interaction picture in the laboratory frame read, analogously to the ones in the ancilla frame,
\begin{equation} \label{eq:IntPictureLab}
    \begin{split} 
        & \hat{\rho}_0^{(L)} = e^{i\hat{H}_{free}^{(L)}\Delta t} \hat{\rho}^{(L)}e^{-i \hat{H}_{free}^{(L)}\Delta t},\\
        & \hat{H}_{int}^{(L)}(\Delta t) = e^{i\hat{H}_{free}^{(L)}\Delta t} \hat{H}_{int}^{(L)}e^{-i \hat{H}_{free}^{(L)}\Delta t},\\
        & i \frac{d \hat{\rho}_0^{(L)}}{dt} = \left[\hat{H}_{int}^{(L)}(\Delta t), \hat{\rho}_0^{(L)} \right],
    \end{split}
\end{equation}
where $\Delta t = t_f - t_i = \gamma_{\pi}^A \Delta \tau$. These equations are satisfied if the QRF transformation in the interaction picture is
\begin{equation}
   \hat{S} =  e^{i \hat{H}_{free}^{(L)}\hat{\gamma}_A \Delta \tau} \hat{S}_S e^{-i \hat{H}_{free}^{(A)}\Delta \tau},
\end{equation}
with $\left[ \hat{H}_{free}^{(L)},\hat{\gamma}_A \right]=0$. This is always verified, as the free Hamiltonian in the laboratory frame cannot depend on the position of the ancilla, given that the free Hamiltonian in the rest frame doesn't and the QRF transformation is a superposition of Lorentz boosts controlled by the velocity of the particle (which is a function of its momentum). The transformation between the rest frame and the laboratory frame in the interaction picture satisfies
\begin{align}
     & \hat{\rho}_0^{(L)} = \hat{S} \hat{\rho}_0^{(A)} \hat{S}^\dagger, \\
     & \hat{H}_{int}^{(L)}(\Delta t) = \hat{\gamma}_A^{-1} \hat{S} \hat{H}_{int}^{(A)}(\Delta \tau) \hat{S}^\dagger,
\end{align}
where the explicit expression of the QRF transformation corresponding to the ``superposition of Lorentz boosts'' in the interaction picture is
\begin{equation} 
   \hat{S} = \mathcal{P}^{(v)}_{LA} \hat{U}_D(\Lambda_{-\hat{p}_L/m_L})\hat{U}_{E}(\Lambda_{-\hat{p}_L/m_L}).
\end{equation}

\bibliography{biblio}

\end{document}